# Modelling imbibition processes in heterogeneous porous media

Si Suo[1], Mingchao Liu[1,2], and Yixiang Gan[1,*]

[1] School of Civil Engineering, The University of Sydney, NSW 2006, Australia

[2] Department of Engineering Mechanics, CNMM & AML, Tsinghua University, Beijing 100084, China

**Abstract:** Imbibition is a commonly encountered multiphase problem in various fields, and exact prediction of imbibition processes is a key issue for better understanding capillary flow in heterogeneous porous media. In this work, a numerical framework for describing imbibition processes in porous media with material heterogeneity is proposed to track the moving wetting front with the help of a partially saturated region at the front vicinity. A new interface treatment, named the interface integral method, is developed here, combined with which the proposed numerical model provides a complete framework for imbibition problems. After validation of the current model with existing experimental results of one-dimensional imbibition, simulations on a series of two-dimensional cases are analysed with the presences of multiple porous phases. The simulations presented here not only demonstrate the suitability of the numerical framework on complex domains but also present its feasibility and potential for further engineering applications involving imbibition in heterogeneous media.

## 1  Introduction

Imbibition as a typical type of capillary flow in porous media is a ubiquitous physical phenomenon, which has a wide range of applications from daily commodities, e.g., napkins and baby diapers, to advanced engineering applications, such as paper-based chromatography (Block, et al., 2016), microfluidics for medical diagnosis (Liu, et al., 2015, Tang, et al., 2017), energy-harvesting devices (Nguyen, et al., 2014), and oil recovery (Morrow, et al., 2001, Rokhforouz, et al., 2017). The research on imbibition phenomena starts from the pioneer work by Lucas (Lucas, 1918) and Washburn (Washburn, 1921) who proposed an analytical model for capillary rise in tubes, known as the Lucas-Washburn equation. Imbibition in porous media shares the same physical principle as that in capillary tubes, i.e., they both are capillary pressure driven liquid flow in essence, though the phenomena are more complicated in porous media. Adequately modelling imbibition processes in porous media is required, in particular when dealing with heterogeneous porous media (Reyssat, et al., 2009), i.e. with spatial distribution and variability of effective properties (Durlofsky, 1991, Warren, et al., 1961).

Motivated by limitations of the Lucas-Washburn equation, a series of modifications have been proposed, to consider the inertia (Quéré, 1997), gravity (Fries, et al., 2008), evaporation (Fries, et al., 2008), and shape and tortuosity of the pore space (Cai, et al., 2014, Cai, et al., 2011). These modified theoretical models replenish the deviation between experimental results and predictions of the original Lucas-Washburn equation, but the corresponding solutions are only available in one-dimensional cases. For higher-dimensional cases, a few analytical solutions for homogeneous porous media of limited geometrical shapes are provided for radial penetration (Conrath, et al., 2010, Liu, et al., 2016), fan-shape membrane (Mendez, et al., 2009), variable-width paper strips (Elizalde, et al., 2015), semi- infinite domain (Xiao, et al., 2012), and fractal porous media (Jin, et al., 2017, Kun-Can, et al., 2017). Recently, the combined effects of geometry and evaporation and gravity are also investigated (Liu, et al., 2018, Xiao, et al., 2018). Moreover, employing the nonlinear Richard's equation, Perez-Cruz *et al.* (Perez-Cruz, et al., 2017) developed a two-dimensional imbibition model that is suitable for arbitrary geometries of homogeneous porous media, i.e., effective properties are spatially independent.

Imbibition in heterogeneous porous media is encountered commonly across different scales. For example, at the macro-scale, layered soils are typical heterogeneous porous media, which are composed of several layers of sediments characterized by grain and pore sizes (Al-Maktoumi, et al., 2015, Zhuang, et al., 2017). Other media may exhibit hierarchical structure across different length scales, including fractured reservoirs and soils containing double porosity (Di Donato, et al., 2007, Lewandowska, et al., 2005). When observed at the mesoscale, concrete materials are typical examples, since the main compositions include aggregates, cements and C-S-H phase, each with distinct permeability. The overall permeability of concretes is determined by this mesoscale structure, which responds for barrier performance, concrete durability (Hall, 1994, Hanžič, et al., 2010, Navarro, et al., 2006), etc. However, earlier work on overall flow properties of heterogeneous porous media (Durlofsky, 1991, Pettersen, 1987, Warren, et al., 1961) focuses on effective permeability instead of detailed imbibition processes, e.g., the evolution of liquid front, whilst recent work are mostly limited to layered configuration (Bal, et al., 2011,

Debbabi, et al., 2017, Ern, et al., 2010, Guerrero-Martínez, et al., 2017, Helmig, et al., 2007, Patel, et al., 2017, Reyssat, et al., 2009, Schneider, et al., 2017). Reyssat *et al*. (Reyssat, et al., 2009) conducted a research on the imbibition process of layered granular media experimentally and theoretically. Guerrero-Martínez *et al*. (Guerrero-Martínez, et al., 2017) carried out a series of numerical simulations on a three-layer porous media and adopted a hyperbolic tangent function to treat the interfaces approximately. In addition, at microscale, Spaid *et al*. (Spaid, et al., 1998) simulated the multicomponent flow in heterogeneous porous media through modified lattice Boltzmann method. Thus, it is necessary to extend the continuum numerical framework to more complex domains, in particular, solving problems with the presence of interfaces among distinct types of porous media.

In this work, a new model for imbibition in heterogeneous porous media is proposed based on the nonlinear Richard's equation, with consideration of spatial distribution of properties. In particular, we develop an efficient and exact method for interface treatment, combined with which the proposed model provides a complete numerical framework for imbibition processes in arbitrarily heterogeneous porous media, including graded and stepwise structures. The developed numerical scheme is verified through comparing with existing experimental results. Finally, imbibition processes in various two-dimensional complex domains are simulated to demonstrate the versatility and suitability of the developed numerical framework.

## 2 Modelling

### 2.1 Liquid front tracking

The capillary imbibition in heterogeneous porous media is much complex than that in homogeneous ones. A generalised model is needed for quantitative prediction of these complex behaviours. The main challenge in the modelling process is to track the moving wetting front (i.e., the boundary between the wetting and non-wetting fluids). In order to be applicable in complex-shaped domains, here we develop a model based on the fact that the liquid content, $\theta$, around the wetting front shows a gradual change rather than a sharp transition depicted in Figure 1(a) and (b), i.e., the value of $\theta$ is between zero and the porosity $\phi$ (Huinink, 2016). Thus, the degree of saturation, $S$, which varies from 0 to 1 as shown in Figure 1(c), is used to describe the partially occupied state of wetting phase in the pore space and taken as the primary unknown,

$$S = \theta/\phi. \tag{1}$$

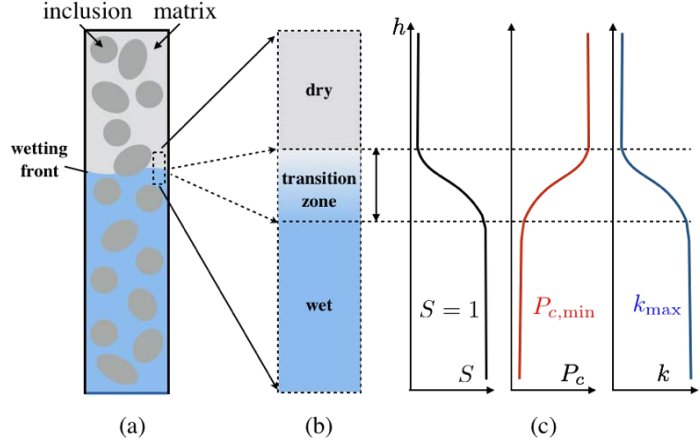

Figure 1. Schematic diagram of non-sharp wetting front: (a) A typical heterogeneous porous medium containing two different sub-domains; (b) Zoom-in view at the transition zone of the wetting front; (c) Profiles of saturation, capillary pressure, and permeability across the transition zone.

Without the external source or sink term, the mass conservation law for capillary imbibition process in porous media reads as follows:

$$\phi \frac{\partial S}{\partial t} = -\nabla \cdot \vec{\mathbf{q}}, \tag{2}$$

where $\vec{\mathbf{q}}$ is the volumetric flux, and according to Darcy's law,

$$\vec{\mathbf{q}} = -\frac{k(\mathbf{x},S)}{\mu}\left(\nabla P_c(\mathbf{x},S) - \rho g \vec{\mathbf{e}}_\mathbf{g}\right), \tag{3}$$

where $\mu$ is the viscosity of the liquid; $\rho g$ represents the gravity-induced pressure gradient and $\vec{\mathbf{e}}_\mathbf{g}$ is a directional unit vector for gravity; $k$ and $P_c$ are the permeability and capillary suction of the matrix material, respectively, and both are the functions of position $\mathbf{x}$ and saturation $S$, as shown in Figure 1(c). The capillary suction, $P_c$, provides driving pressure for imbibition. Thus, Eq. (2) can be rewritten as

$$\frac{\partial S}{\partial t} = \nabla \cdot \frac{k(\mathbf{x},S)}{\phi \cdot \mu}\left(\nabla P_c(\mathbf{x},S) - \rho g \vec{\mathbf{e}}_\mathbf{g}\right). \tag{4}$$

Different forms of $k$ and $P_c$ can be found in previous studies on various types of porous media. Here, Brooks-Corey relations (Brooks, et al., 1964) are adopted, as an example, to describe the saturation-dependent $k(\mathbf{x},S)$ and $P_c(\mathbf{x},S)$ as

$$k(\mathbf{x},S) = k_{max}(\mathbf{x}) \cdot S^{2/\lambda+3}, \tag{5}$$

$$P_c(\mathbf{x},S) = P_{c,min}(\mathbf{x}) \cdot S^{-1/\lambda}, \tag{6}$$

where $\lambda$ is the imbibition index by which the width of the transition zone is controlled and more discussion regarding $\lambda$ can be found in Ref (Perez-Cruz, et al., 2017). Note that here the format of Brooks-Corey relations is used for describing the liquid front, and $\lambda$ is indirectly related to the intrinsic material properties. The discussion regarding this correlation can be found later in Section 4. Thus, a smaller $\lambda$ value results in a sharper transition between wetting and non-wetting fluids, while a larger value will smear the liquid front. The maximum value of the permeability, $k_{max}$, is proportional to the square of characteristic pore size, $d^2$, and a certain expression of the porosity $\phi$ according to the type of porous media. For granular materials, $k_{max}$ can be expressed as (Bear, 2013)

$$k_{max} = \frac{d^2}{180} \cdot \frac{\phi^3}{(1-\phi)^2}, \tag{7}$$

where $d$ is the average grain size. For other types of porous materials, such as fibre composite materials and cementations materials, $k_{max}$ can be estimated by (Huinink, 2016)

$$k_{max} = \frac{1}{8} \cdot \phi \cdot d^2. \tag{8}$$

Moreover, $P_{c,min}$, the minimum value of the capillary pressure, generally has the formula as follows:

$$P_{c,min} = \frac{c^* \cdot \gamma}{d}, \tag{9}$$

where $\gamma$ is the surface tension and $c^*$ is a coefficient related to the contact angle.

For heterogeneous porous media, we consider here the scenario that the characteristic pore size, $d$, varies over the domain as $d(\mathbf{x})$. Substituting Eq. (5) and Eq. (6) into Eq. (4) and neglecting the contribution of gravity, the governing equation can be further derived as

$$\frac{\partial S}{\partial t} = \nabla \cdot \left( \frac{k_{max}(\mathbf{x}) \cdot P_{c,min}(\mathbf{x})}{\lambda \cdot \phi \cdot \mu} \cdot S^{1/\lambda+2} \cdot \nabla S - \frac{k_{max}(\mathbf{x})}{\phi \cdot \mu} \cdot S^{1/\lambda+3} \cdot \nabla P_{c,min}(\mathbf{x}) \right). \tag{10}$$

The right hand of Eq. (10) is divided into two terms, in which the first term, centred with $\nabla S$, describes liquid transfer induced by saturation gradient, i.e., naturally it can be regarded as a diffusion process. Since $P_{c,min}(\mathbf{x})$ is related to the property of matrix materials as shown in Eq. (9), the second term is an additional flux induced by the non-homogeneity of matrix materials, while it is spontaneously eliminated for homogenous matrix materials.

## 2.2 Interface treatment

Depending on the actual spatial distribution of the material properties, Eq. (10) can be applied to provide the evolution law for the local degree of saturation. However, for some special cases with the presence of interfaces, e.g., in layered porous materials (Guerrero-Martínez, et al., 2017, Reyssat, et al., 2009), singularity needs to be handled in the numerical solutions. This singularity appears when differentiating a step function at the interface where the corresponding material property has a sharp change. Conventional solutions to deal with this type of problems include (1) smoothing the step function and (2) requiring special mesh treatment at the vicinity of the interface, which are difficult for higher-dimensional cases and likely result in mesh-dependent predictions for imbibition processes. In this work, based on the weak form of Eq. (10), we propose a new interface treatment approach, named the *interface integral method*, which is not only suitable for simple one-dimensional cases but also for the higher-dimensional interfaces of complex geometry.

A typical calculation region for a multi-media system is presented in Figure 2, and by introducing a test function, $S^*(\mathbf{x},t)$, Eq. (4) without gravity can be developed into the corresponding equivalent integral weak form as follows:

$$\int_\Omega \frac{\partial S}{\partial t} \cdot S^* \cdot d\Omega = -\int_\Gamma S^* \cdot \frac{k(\mathbf{x},S)}{\phi \cdot \mu} \nabla P_c(\mathbf{x},S) \cdot \vec{n} \cdot d\Gamma + \int_\Omega \nabla S^* \cdot \frac{k(\mathbf{x},S)}{\phi \cdot \mu} \nabla P_c(\mathbf{x},S) \cdot d\Omega, \quad (11)$$

where the meanings of $\Omega$, $\Gamma$, and $\vec{n}$ can be referred in Figure 2.

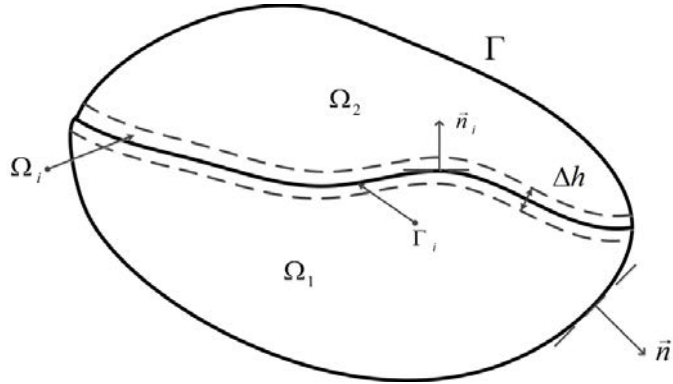

Figure 2. Schematic of a typical heterogeneous domain, containing an exterior boundary $\Gamma$ whose normal vector is $\vec{n}$ and two domains $\Omega_1$ and $\Omega_2$ that share a mutual interface $\Gamma_i$ whose normal vector is $\vec{n}_i$.

Generally, Eq. (11) can be implemented using a standard FEM scheme (Zienkiewicz, et al., 1977) for imbibition problems in certain porous media if the material properties vary continuously. However, a further improvement is needed since the material property has a sharp change at the interface. Due to this discontinuity, an interfacial transition region $\Omega_i$ with infinitesimal thickness $\Delta h$ is built at the vicinity of the interface, as shown in Figure 2. Subsequently, the integral area of the second term on the right hand in Eq. (11) can be separated into three domains, i.e., $\Omega_1$, $\Omega_2$, and $\Omega_i$,

$$\int_\Omega \nabla S^* \cdot \frac{k(\mathbf{x},S)}{\phi \cdot \mu} \nabla P_c(\mathbf{x},S) \cdot d\Omega = \int_{\Omega_1+\Omega_2} \nabla S^* \cdot \frac{k(\mathbf{x},S)}{\phi \cdot \mu} \nabla P_c(\mathbf{x},S) \cdot d\Omega \\ + \int_{\Omega_i} \nabla S^* \cdot \frac{k(\mathbf{x},S)}{\phi \cdot \mu} \nabla P_c(\mathbf{x},S) \cdot d\Omega \qquad (12)$$

The integral expression on $\Omega_i$, named as the interface integral term $I_i$, can be rewritten equivalently as follows considering that $\Delta h$ is an infinitely small quantity,

$$I_i = \int_{\Omega_i} \nabla S^* \cdot \frac{k(\mathbf{x},S)}{\phi \cdot \mu} \nabla P_c(\mathbf{x},S) \cdot d\Omega = \lim_{\Delta h \to 0} \Delta h \cdot \int_{\Gamma_i} \nabla S^* \cdot \frac{k(\mathbf{x},S)}{\phi \cdot \mu} \nabla P_c(\mathbf{x},S) \cdot d\Gamma. \qquad (13)$$

Then, through substituting Eq. (5) and (6), Eq. (13) can be divided into two parts as

$$I_i = \lim_{\Delta h \to 0} \Delta h \cdot \int_{\Gamma_i} \nabla S^* \cdot \frac{k_{\max}(\mathbf{x}) \cdot P_{c,\min}(\mathbf{x})}{\lambda \cdot \phi \cdot \mu} \cdot S^{1/\lambda+2} \cdot \nabla S \cdot d\Gamma \\ + \lim_{\Delta h \to 0} \Delta h \cdot \int_{\Gamma_i} \nabla S^* \cdot \frac{k_{\max}(\mathbf{x})}{\phi \cdot \mu} \cdot S^{1/\lambda+3} \cdot \nabla P_{c,\min}(\mathbf{x}) \cdot d\Gamma \qquad (14)$$

As for the first term in Eq. (14), $\nabla S$ has finite value considering the continuity of $S$ around the interface and so do the values of $k_{\max}$ and $P_{c,\min}$ for a given porous medium. Therefore, the integral in the first term on the right hand side is a limited quantity and it becomes zero when $\Delta h$ approaches zero. As for the second term, $\nabla P_{c,\min}(\mathbf{x})$ along the interface can be expressed as

$$\nabla P_{c,\min} = \frac{P_{c,\min}^+ - P_{c,\min}^-}{\Delta h} \cdot \vec{n}_i, \qquad (15)$$

where $P_{c,\min}^+$ and $P_{c,\min}^-$ are $P_{c,\min}$ within the domains, $\Omega_2$ and $\Omega_1$, respectively.

By substituting Eq. (15) into Eq. (14), the integral term $I_i$ is simplified as

$$I_i = \int_{\Gamma_i} \nabla S^* \cdot \frac{\bar{k}_{\max}}{\phi \cdot \mu} \cdot S^{1/\lambda+3} \cdot \left( P_{c,\min}^+ - P_{c,\min}^- \right) \cdot \vec{n}_i \cdot d\Gamma, \qquad (16)$$

where $\bar{k}_{\max}$ is here taken as the mean value of $k_{\max}$ within the domains,

representing the upper bound (Hashin, et al., 1963).

From the above, the generalised weak form of the heterogeneous system is finally proposed as follows,

$$\int_\Omega \frac{\partial S}{\partial t} \cdot S^* \cdot d\Omega = -\int_\Gamma S^* \cdot \frac{k(\mathbf{x},S)}{\phi \cdot \mu} \nabla P_c(\mathbf{x},S) \cdot \vec{n} \cdot d\Gamma$$
$$+ \int_\Omega \nabla S^* \cdot \frac{k(\mathbf{x},S)}{\phi \cdot \mu} \nabla P_c(\mathbf{x},S) \cdot d\Omega \qquad (17)$$
$$+ \int_{\Gamma_i} \nabla S^* \cdot \frac{k_{\max}(\mathbf{x})}{\phi \cdot \mu} \cdot S^{1/\lambda+3} \cdot \left(P_{c,\min}^+ - P_{c,\min}^-\right) \cdot \vec{n}_i \cdot d\Gamma$$

Compared with the Eq. (11), the weak form Eq. (17) for heterogeneous systems contains an additional interface integral term $I_i$ induced by the presence of sharp changes of material properties at interfaces. Combined with various discretisation methods, Eq. (17) can be used in different numerical frameworks of various mesh-based and meshless methods, such as FEM and SPH.

### 2.3 Numerical realization

In this work, COMSOL Multiphysics®, a partial differential equation solver based on finite element method, was adopted to obtain the numerical solution of Eq. (10). Special treatment has been introduced in WEAK CONTRIBUTION module to implement the weak form for interface regimes.

To summarise, we develop a complete and generalised model for imbibition processes, and this proposed model is applicable to various types of homogeneous and heterogeneous porous media when $k_{\max}(\mathbf{x})$ and $P_{c,\min}(\mathbf{x})$ are defined accordingly. Moreover, an interface integral method is also proposed to consider the presence of interfaces in the heterogeneous porous media. In order to verify the practicability of the proposed model and present its potential applications, several numerical cases are provided in the following sections.

## 3 Numerical validations and examples

In this section, we validate the proposed imbibition model against available experimental data and further test more generalised numerical examples. First, in Sections 3.1 and 3.2, the model is validated by two one-dimensional experimental cases from Ref (Reyssat, et al., 2009), where the flow direction was predefined and aligned with the orientation of the changing properties. Subsequently, typical examples of two-dimensional cases are used to demonstrate the potential of this numerical scheme. Several parameters regarding material properties in Eq. (10) are shown as follows: $\phi = 0.38$, $\gamma = 0.02$ N/m, $\mu = 0.01$ Pa·s, $c^* = 0.02$ and $\lambda = 0.1$. Here these are typical for silicon oil, as an example, but other settings are also feasible.

**3.1 Granular media with permeability gradients**

In the experimental work (Reyssat, et al., 2009), two types of granular media, i.e., with permeability gradient and two-layered structures, were prepared and tested for imbibition of silicon oil. For the former, beads with different sizes arranged in a horizontal tube form the porous media with controlled permeability gradient, as shown in Figure 3(a), and this case can be simplified as a one-dimensional calculation model by assuming that the bead column is slim enough that the non-uniformity of pressure in the tube cross section can be neglected. As for boundary conditions, the end immersed in the liquid reservoir is represented by a Dirichlet boundary condition $S = 1$ and the other end is set as a zero-flux Neumann boundary condition, i.e., $\vec{q} \cdot \vec{n} = 0$, in Figure 3(b). The linear change of $k_{max}$ was constructed in the experiments, i.e., $k_{max} = k_0 + \beta \cdot x$, where $k_0$ is the permeability at the opening of the tube and $\beta$ is the permeability gradient along the x-direction. From Eq. (7), we can have the relation for the grain size as $d(x) = \sqrt{\dfrac{180 \cdot (1-\phi)^2 \cdot (k_0 + \beta \cdot x)}{\phi^3}}$, and then substitute it into Eq. (9). The numerical solutions for Eq. (10) are presented in Figure 3(c) and compared with corresponding experimental results (Reyssat, et al., 2009).

As shown in Figure 3(c), the numerical results of the proposed model are in good agreement with experimental measurements of the liquid front for granular materials with different permeability gradients ($\beta$), and specifically the relative errors of these cases are less than 4%. Furthermore, the numerical solutions of Eq. (10) capture the intrinsic imbibition behaviour in gradually-changed porous media, i.e., (1) a log-log linear relationship between the position of liquid front and imbibition time, and (2) a significantly higher advancing speed of the liquid front in granular medium with decreasing grain sizes than the one with increasing grain sizes along the flow direction. For the latter, this demonstrates the significance of the second term on the left hand side of Eq. (10) for graded porous media.

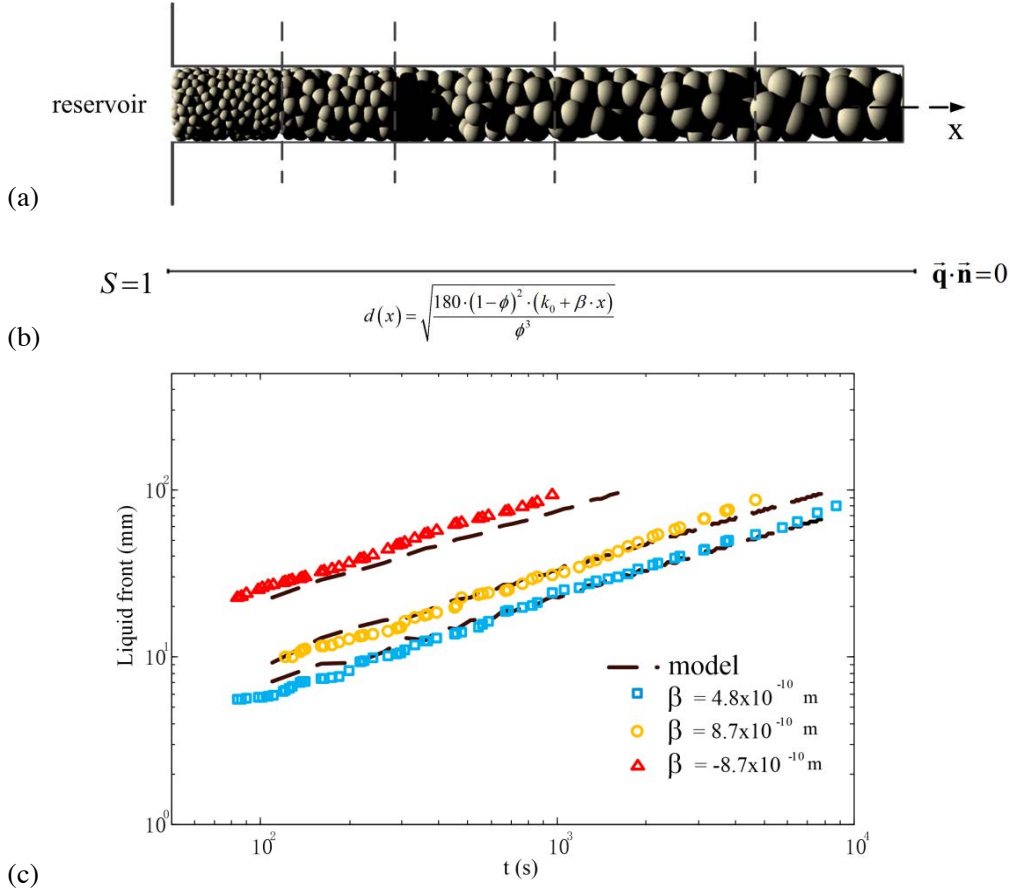

Figure 3. (a) Schematic of a granule-filled tube including several layers with varying the bead size (*d*) to mimic the linearly changing local permeability along *x*-axis; (b) One-dimension simulation model including boundary conditions and the bead size as a function of length; (c) Position of the imbibition front as a function of time for the system with different permeability gradients, including $\beta$ = 4.8x10$^{-10}$ m (□), $\beta$ = 8.7x10$^{-10}$ m (○) and $\beta$ = -8.7x10$^{-10}$ m (△), compared with simulation results (lines).

### 3.2 Two-layer granular materials

Imbibition processes into two-layer granular materials are modelled here for validating the proposed numerical treatment at the interface. Two-layer systems were formed by arranging two types of beads with different sizes in a tube (Reyssat, et al., 2009). The one-dimensional numerical model with the same boundary conditions as mentioned in *Section 3.1* can also simulate this case, as shown in Figure 4(a) and (b). In particular, the interfacial treatment has been implemented here to handle the sharp transition between these two layers, to resolve numerical singularities due to discontinuities of the derivatives.

It can be seen from Figure 4(c) that simulation results also exhibit decent agreement with the experimentally measured liquid front motion and the maximum relative error of these cases is less than 14%. In Ref (Reyssat, et al., 2009), the authors also provided a piece-wise theoretical solution for two-layered cases, in which the advancing speed of liquid front drops since the flux continuity condition at the interface was ignored, and thus violating the mass conservation law. However, for the current model, the continuity condition is satisfied at the interface automatically, via Eq. (10), so that both the position and advancing speed of the liquid front maintain continuous at the interface.

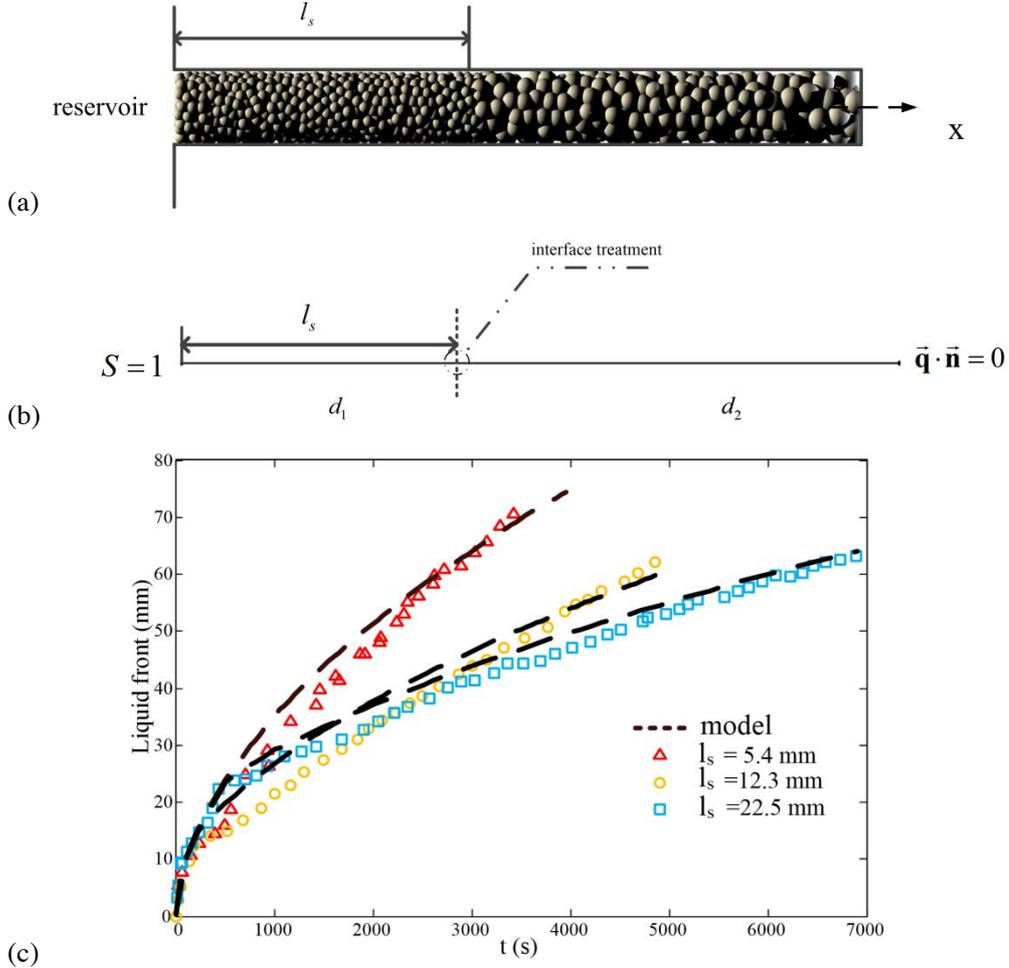

(a)

(b)

(c)

Figure 4. (a) Schematic of a granule-filled tube including two layers with a jump between regions consisting of small beads ($d_1$) and large beads ($d_2$) at a distance $l_s$; (b) One-dimension simulation model including boundary conditions and treatment of the interface between two layers; (c) Position of the imbibition front as a function of time for the system with $d_1$ = 41 μm and $d_2$ = 196 μm, and three cases of different $l_s$ were included, i.e., $l_s$ = 5.4 mm (△), 12.3 mm (○) and 22.5 mm (□), compared with simulation results (lines).

Through the numerical examples and their comparisons to the experimental data in previous and current sections, it is shown that the kinetics of the wetting front is associated with the actual spatial distributions of the measureable material properties, e.g., pore (grain) size. The numerical cases presented here not only verified the proposed simulation scheme including governing equations and interface treatments, but also lay the root for more complex applications since interface issues are ubiquitous in various fields.

### 3.3 Regions with oblique interfaces

After validating the case with the presence of interfaces, we show a few general examples here to demonstrate the potential of this proposed numerical framework. Oblique interfaces are commonly encountered in heterogeneous porous media, in particular, in geology and sedimentology. As shown in Figure 5, a rectangular region composed of two types of porous materials that are separated by a 45° oblique interface forms a typical two-dimensional case. The general interface treatment is also imposed on oblique interfaces.

Figure 5 (a) and (b) show different spatial arrangements of porous regions, PM1 with characteristic pore size $d_1$ = 41 μm and PM2 with $d_2$ = 196 μm, which generate different imbibition patterns. The boundary conditions were prescribed as follows, (1) Dirichlet boundary condition, $S=1$, is applied at the bottom edge, and (2) zero-flux conditions are imposed on the other three edges. The gravity term is ignored in this example. First of all, the soaking speed of Case I (the smaller pore size region, PM1, on the top) is significantly faster than that of Case II (the larger pore size region, PM2, on the top) through comparing the time it takes to form final configures, i.e., the whole imbibition rate is determined by the local property of porous media close to the reservoir. As for the liquid front across the interface, in Case I, a "drag effect" was observed for the imbibition process when the liquid front invades from the region with a larger pore size (PM2) to PM1, and tangent lines of liquid fronts at the boundary of two regions form a convex angle (marked in yellow in Figure 4). Whilst reversely in Case II the imbibition in PM1 region impede that in PM2 and thus a concave angle (marked in green) is formed at the interface, i.e., a "pull effect".

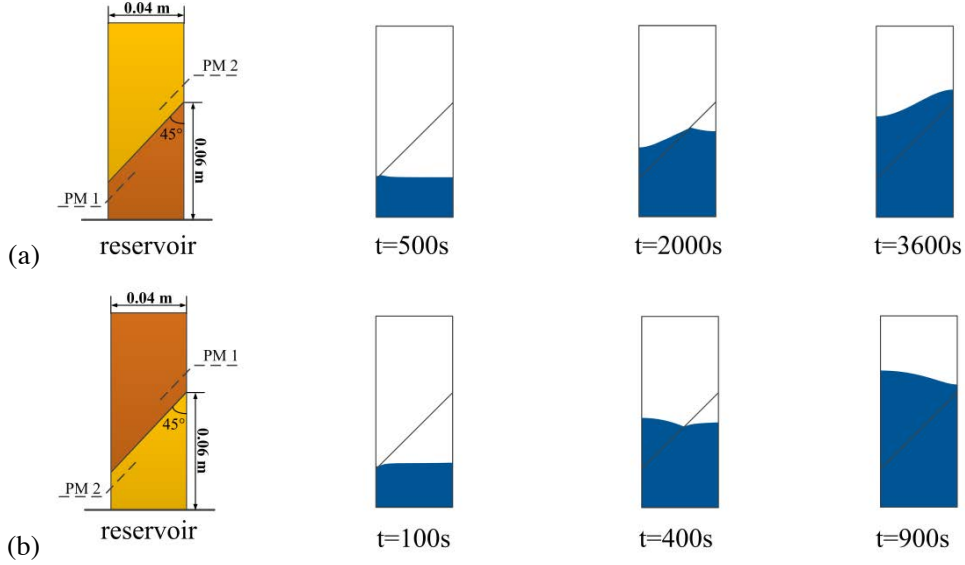

Figure 5. A rectangular domain with two types of porous materials PM1 (■) with smaller pore size and PM2 (■) with larger pore size separated by an oblique sharp interface. Time lapses for (a) Case I: smaller pore size region, PM2, on the top; (b) Case II: larger pore size region, PM1, on the top.

## 3.4 Regions with inclusions

Inclusions in heterogeneous materials are commonly observed, such as concrete (aggregates and cement paste) and fibre bundle composites (Hall, 1994, Hanžič, et al., 2010, Navarro, et al., 2006, Spaid, et al., 1998), where both matrix and inclusion phases can be treated as porous media with different material properties. Such differences between phases can influence the imbibition process. Here, we show relevant numerical examples to illustrate the imbibition tracking at the inclusion scale, which can provide rich information for modelling such processes in these composite materials. In Figure 6(a) and (b), an elementary volume of the inclusion problem is established, in which a circular inclusion, and thus a curved interface, is included. The material parameters and boundary conditions are the same as that of the above case presented in Section 3.3, e.g., PM1 with characteristic pore size $d_1$ = 41 μm and PM2 with $d_2$ = 196 μm. Case I and II are arranged with different spatial allocation of the material properties, and the simulations show distinct imbibition patterns. For Case I, the liquid front in the outer domain (matrix) is always higher than that in the inner domain (inclusion) and the liquid front in the inclusion advances in a downward concave shape due to the drag effect mentioned above. For Case II, on the contrary, the liquid front in the inner domain goes gradually higher than that in the outer domain and an upward convex liquid front is caused by the pull effect introduced by the inclusion phase.

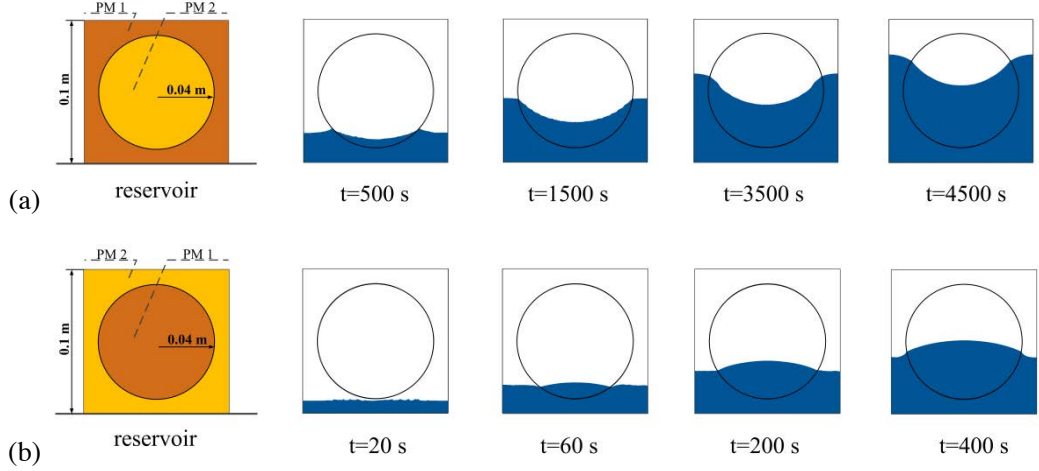

Figure 6. A square porous matrix (side length is 0.1 m) contains a circular inclusion (radius is 0.04 m) where two types of porous materials PM1 (▬) with a smaller pore size and PM2 (▬) with a larger pore size are considered. Time lapses for (a) Case I: PM1 for the matrix while PM2 for the inclusion; (b) Case II: PM2 for the matrix while PM1 for the inclusion.

## 4  Typical heterogeneous porous media

In order to investigate the imbibition property in typical heterogeneous porous media, a macroscale model, as shown in Figure 7, is taken as a representative elementary volume (REV) in which circular inclusions distribute randomly in the matrix, and periodic boundary condition is applied on both right and left side.

The simplest situation for one-dimensional imbibition can be modelled by Lucas-Washburn's equation, i.e., dynamics in capillary tubes (Lucas, 1918, Washburn, 1921),

$$l(t) = \sqrt{2 \cdot D \cdot t}, \qquad (18)$$

where $l$ is position of liquid front; $D$ is diffusive coefficient and generally $D = k \cdot P_c / \mu$. For convenience in discussing imbibition in heterogeneous porous media, we introduce effective liquid front $l_{eff}$,

$$l_{eff} = A_w / b, \qquad (19)$$

where $A_w$ is wetting area, i.e., $A_w = \int_\Omega (S - S_0) \cdot d\Omega$, and $b$ is the width of the selected REV. In Figure 7(a) and (b), cases with different pore size ratio, defined as

$$d_1 = \gamma \cdot d_0, \qquad (20)$$

where $d_0$ and $d_1$ are the characteristic pore sizes of the matrix and inclusion,

respectively, and $\gamma$ is the pore size ratio. The results show that the effective liquid front $l_{eff}$ as a function of time $t$ is similar to the predictions of Lucas-Washburn's equation.

Besides, pore size ratio of the particles and matrix casts a significant influence on imbibition efficiency for a given heterogeneous porous media. It is found that the case with the pore size ratio of one ($\gamma = 1$, indicating a homogenous medium) shows the slowest imbibition process. This result seems to be counter-intuitive, i.e., the presence of the inclusion phase, whether its intrinsic diffusivity is higher or lower than the one of matrix, always accelerates the overall imbibition process. To depict this effect, an effective diffusive coefficient $D_{eff}$ is introduced here and it is obtained through fitting simulation data to the original Lucas-Washburn equation, as Eq. (18). For homogenous porous media, the intrinsic diffusive coefficient can be defined as

$$D = \psi(\lambda) \cdot \left(k_{max} \cdot P_{c,min}\right) / (\phi \cdot \mu), \qquad (21)$$

where $\psi$ is a coefficient relate to the imbibition index $\lambda$, introduced in Section 2, and it is calculated through an averaging process inside the transition zone by

$$\psi(\lambda) = \int_0^\infty \frac{S(\xi,\lambda) - S_{min}}{S_{max} - S_{min}} \cdot d\xi \qquad (22)$$

where $\xi$ is a scaling variable regarding position and time, and more details on this derivation can be referred to Ref (Huinink, 2016). Through the numerical integration, $\psi$ is around 0.77 for the case of $\lambda = 0.1$, and it is found the actual value of $\psi$ is insensitive to $\lambda$ around 0.1.

Furthermore, for two-component porous media shown in Figure 7, the effective diffusive coefficient is also determined by two parts, i.e., permeability related and capillary suction related parts. In regard to the former, the bounds of effective maximum permeability are evaluated according to the classic theory of effective properties of two-component materials (Böttcher, et al., 1978, Zimmerman, 1989), i.e., parallel model and series model; for the latter, the weighted average of $P_{c,min}$ of two components is adopted here as the effective minimum capillary pressure due to assumption that the inclusions distribute uniformly in the matrix. Combining these two parts, the upper bound and the lower bound of effective diffusive coefficient are evaluated here respectively,

$$D_{upper} = \psi(\lambda) \cdot \left(v_0 \cdot k_{max}^0 + v_1 \cdot k_{max}^1\right) \cdot \left(v_0 \cdot P_{c,min}^0 + v_1 \cdot P_{c,min}^1\right), \qquad (23)$$

$$D_{lower} = \psi(\lambda) \cdot \frac{k_{max}^0 \cdot k_{max}^1}{v_1 \cdot k_{max}^0 + v_0 \cdot k_{max}^1} \cdot \left(v_0 \cdot P_{c,min}^0 + v_1 \cdot P_{c,min}^1\right), \qquad (24)$$

where $v_0$, $k_{max}^0$ and $P_{c,min}^0$, and $v_1$, $k_{max}^1$ and $P_{c,min}^1$ are volume fractions, maximum permeability, and minimum capillary suction of matrix and inclusions, respectively. As shown in Figure 8, the effective diffusivity obtained from numerical simulations with $\gamma$ ranging from 0.2 to 5.0 forms a U-shaped curve with a minimum value around $\gamma = 1$. When compared with the upper and lower bound predictions, the cases of pore size ratio smaller than one, $D_{eff}$ tend to approximate the upper bound, while the lower bound gives a better estimation of $D_{eff}$ when pore size ratio is larger than one. Thus, this work can be extended as a numerical tool for solving inverse problems and predicting the effective diffusivity of composite materials, utilising the spontaneous imbibition processes.

# 5 Discussion

Based on the proposed numerical framework, imbibition processes are effectively simulated in heterogeneous porous media ranging from one-dimensional to higher-dimensional cases. In this framework, the liquid front is captured by implicitly imposing a narrow transition zone, and a special interface treatment has also been proposed and implemented for cases with the presence of material interfaces.

Generally, for layered porous media, the interfaces cannot be ignored and it brings certain effects on the imbibition process, such as additional flow resistance shown in Section 3.2. Also, the configuration shown in Section 3.3 can be adopted to simulate the groundwater flow and the resultant effective stress states for certain slopes where the interfaces between layered soils may be non-parallel to the underwater level (Al-Maktoumi, et al., 2015, Zhuang, et al., 2017). In fact, the interface issue often occurs in various types of porous media. For concrete, as a typical example, aggregates, cements and C-S-H phases have typical porous media interfaces due to their distinct permeability, and the numerical cases shown in Section 3.4 can work as a simplified REV model to analyse barrier performance of concrete structures (Hall, 1994, Hanžič, et al., 2010, Navarro, et al., 2006). Besides interface issues, the porous media with gradually-changed permeability also presents special flow pattern and liquid front dynamics, as shown in Section 3.1, which can guide the design of paper-based microfluidic chips (Block, et al., 2016, Nguyen, et al., 2014, Tang, et al., 2017). The example presented in Section 4 is a typical heterogeneous porous medium, which can be used to analyse soil containing two sub-domains characterised by contrasted pore sizes (Lewandowska, et al., 2005), and the present approach provides rich information from the microstructural heterogeneity. Based on the analysis in Section 4, a microstructure informed prediction is developed, i.e., the effective diffusive parameter of the whole heterogeneous structure can be estimated under the proposed framework, as an analogue to the spontaneous imbibition experiments of fibrous composites (Cai, et al., 2012) and carbonate rocks (Alyafei, et

al., 2018).

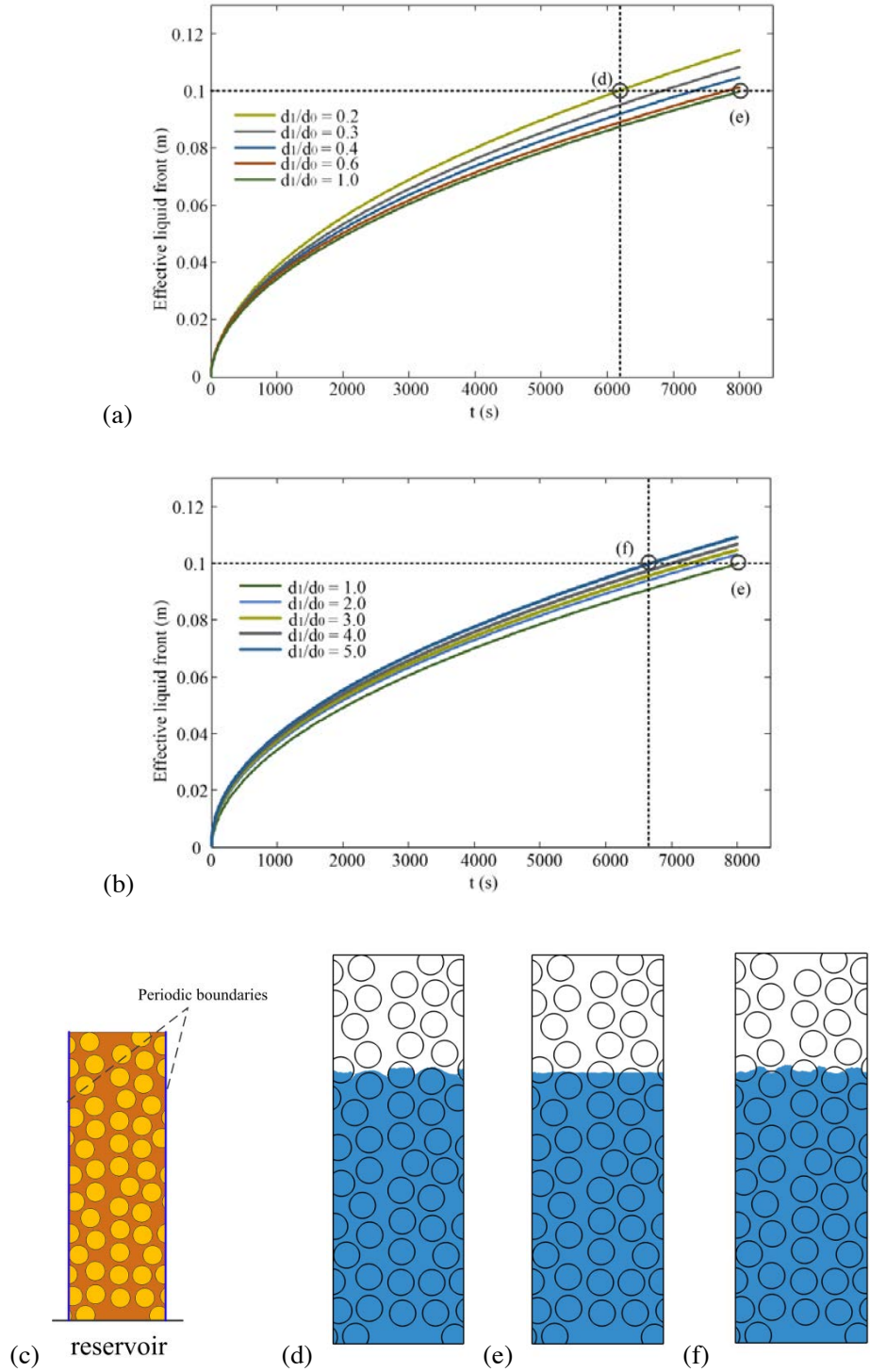

Figure 7. A macroscale porous matrix (width is 0.05 m) with randomly-distributed circular inclusions (radius is 0.005 m), where two types of porous materials are included: matrix (▬) with a smaller pore size $d_0$ = 50 μm and inclusions (▬) with a pore size $d_1 = \gamma \cdot d_0$ ($\gamma$ = 0.2, 0.3, 0.4, 0.6, 1.0, 2.0, 3.0, 4.0, and 5.0). (a-b) Effective liquid front position as a function of time; (c) Model schematic; (d-f) Wetting profiles with pore size ratio is 0.2, 1, and 5, respectively.

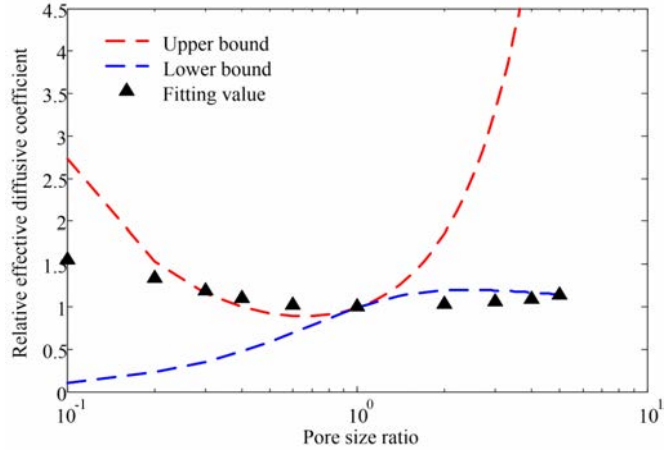

Figure 8. Relative effective diffusive coefficient $D_{eff}/D_0$ vs. pore size ratio of the matrix and the inclusions, where $D_0$ is the diffusive parameter when pore size ratio equals one.

# 6 Conclusion

In this work, we develop a comprehensive numerical framework to solve imbibition problems in heterogeneous porous media. Detailed wetting front dynamics has been captured by introducing the partially saturated transition zone at the front vicinity. To extend the method to more complex domains, special interface treatments, i.e., the interface integral method, have also been proposed and implemented. To validate the proposed numerical model, simulations are compared with experimental results of layered porous media and good agreements are achieved. The dynamics of wetting front is found to be associated with the spatial distributions of the material properties, e.g., pore size. Furthermore, combined with the proposed interface treatment, the model is capable of predicting imbibition processes in porous media with arbitrary topology. The proposed framework in this paper contributes to discover the underlying physics behind imbibition processes in heterogeneous materials and to improve quantitative and microstructure informed perditions for capillary flow related phenomena in various fields.

## Acknowledgements

This work was financially supported by Australian Research Council (Projects DP170102886) and The University of Sydney SOAR Fellowship.